\documentclass[%
 reprint,
 amsmath,amssymb,
 aps,
]{revtex4-2}

\usepackage{graphicx}
\usepackage{dcolumn}
\usepackage{bm}
\usepackage{color}
\usepackage{multirow}

\begin{document}

\preprint{APS/123-QED}

\title{Electron Acceleration by the Axisymmetric TE$_{011}$ Mode in a Slowly Varying External Magnetic Field\\}
 
\author{Oswaldo Otero}
\email{oswaldoterolarte@gmail.com}
\author{Jes\'us E. L\'opez}
\author{P. Tsygankov}%
\author{Carlos J. P\'aez-Gonz\'alez}%
\author{E. A. Orozco}%
\email{eaorozco@uis.edu.co}
\affiliation{%
 Universidad Industrial de Santander. A.A. 678 Bucaramanga, Colombia.\\
}%




\date{\today}

\begin{abstract}

This study investigates the autoresonant acceleration of electrons using the GYRAC mechanism in a cylindrical cavity excited in the \( \mathrm{TE}_{011} \) microwave mode, under a slowly increasing external magnetic field. The acceleration process is driven by the interaction between electrons and the right-hand circularly polarized (RHP) component of the electric field, maintained in phase through electron cyclotron resonance (ECR). A simplified single-particle model and numerical simulations based on the relativistic Newton–Lorentz equation were employed to evaluate particle dynamics under different values of the magnetic field growth parameter \( \alpha \).

The results demonstrate that efficient trapping and energy transfer can be achieved for appropriate values of \( \alpha \), and that the spatial non-uniformity of the \( \mathrm{TE}_{011} \) mode introduces critical sensitivity to initial particle positioning. The study further identifies a ring-shaped region in which electrons are consistently captured in the GYRAC regime, and shows how the evolution of the Larmor radius, phase shift, and energy distribution depends on the electromagnetic configuration.

Simulations involving disk-like and ring-like electron clouds reveal the fraction of captured, escaped, and uncaptured particles, confirming that precise control of the magnetic field variation is essential to optimize the efficiency of the acceleration process. These findings provide a basis for future self-consistent plasma simulations and support the development of compact radiation sources based on GYRAC–\( \mathrm{TE}_{011} \) acceleration.

\end{abstract}

\maketitle


\section{\label{sec:Intro}INTRODUCTION}

Since 1962, Kolomenskii and Davydovskii initiated the study of the interaction between a transverse electromagnetic wave propagating along a homogeneous magnetic field and the motion of a charged particle \cite{Kolomenskii1963, davydovski1962possibility}. This phenomenon, known as autoresonance, involves maintaining the equality between the frequency of the electromagnetic wave and the electron cyclotron frequency. An important characteristic of autoresonant interaction is that the decrease in the electron cyclotron frequency, which occurs as the electron gains energy, can be precisely offset by the Doppler shift \cite{Kolomenskii1963, davydovski1962possibility}. This compensation enables sustained interaction with the wave field and forms the basis of several theoretical and experimental studies \cite{jory1968charged, milant2013cyclotron,loeb1986autoresonance, salamin2015feasibility}. For example, Jory and Trivelpiece investigated electron acceleration by means of both linearly and circularly polarized homogeneous plane waves, as well as transverse electric modes ($\mathrm{TE}_{11}$) in circular waveguides, all under the influence of a non-uniform but static axial magnetic field \cite{jory1968charged}. Based on this, Shpitalnik et al. proposed an innovative microwave autoresonant accelerator (AMA), which uses a circularly polarized $\mathrm{TE}_{11}$ mode and a spatially variable magnetostatic field to achieve continuous electron acceleration \cite{shpitalnik1991autoresonance}. Further developments explored acceleration schemes within cavity resonators of circular cross-section operating in the $\mathrm{TE}_{111}$ mode and immersed in a uniform axial magnetic field \cite{mcdermott1985production}. However, due to the relativistic variation of the cyclotron frequency with electron energy, it is not possible to maintain exact resonance throughout the entire acceleration path. To address this, the magnetic field is typically optimized so that the radiofrequency (RF) signal remains in approximate resonance with the average cyclotron frequency experienced by the electron during its trajectory.\\
\noindent A. Neishtadt and A. Timofeev demonstrated that autoresonant acceleration can significantly contribute to electron cyclotron heating in plasmas, as electrons are able to sustain the ECR condition while moving along magnetic field lines into regions of increasing magnetic strength \cite{neishtadt1987autoresonance}. This mechanism, known as spatial autoresonance, has provided a foundation for the design of accelerator cavities for electron beams \cite{dugar2009cyclotron, dugar2017compact,otero2019numerical, velazco1993discussion, velazco2003development, velazco2016novel}. In such systems, the magnetostatic field profile is tailored to maintain resonance between the microwave and cyclotron frequencies along the particle trajectory.  Theoretical investigations have explored Spatial Autoresonance Acceleration (SARA) using transverse electric modes $\mathrm{TE}_{11p}$ with mode indices $p=1,2,3$ \cite{dugar2009cyclotron,otero2019numerical}, and an X-ray source based on the SARA principle has been patented \cite{dugar2017compact}. More recently, Velazco et al. proposed a novel compact accelerator concept based on spatial autoresonance and employing the $\mathrm{TM}_{110}$ mode, referred to as the Rotating-Wave Accelerator (RWA) \cite{velazco2016novel,velazco2003development} whose nonlinear dynamics was recently studied \cite{orozco2024electron}.\\
\noindent In the case of gyroresonant acceleration (GYRAC), an external magnetic field that increases slowly over time is used to sustain the electron cyclotron resonance (ECR) condition \cite{golovanivsky1980autoresonant,golovanivsky1982gyrac,golovanivsky1983gyromagnetic,gal1989gyrac}. This mechanism has demonstrated significant potential in several applications, such as the electron cyclotron resonance ion proton accelerator (ECR-IPAC) for cancer therapy \cite{inoue2014design}, and in the control of relativistic electron bunches in plasmas \cite{andreev2017gyromagnetic}.\\
The GYRAC mechanism has been extensively studied, both theoretically and experimentally, using microwave fields in cylindrical cavities operating in the $\mathrm{TE}_{111}$ mode. These investigations have demonstrated the effectiveness of gyromagnetic autoresonance in enabling controlled electron acceleration within resonant structures. Despite the considerable progress achieved over the past decades, GYRAC remains an active area of research, driven by its potential in the development of compact and efficient electron accelerators \cite{andreev2020generation, andreev2021autoresonance}.

The present work investigates electron acceleration using the axisymmetric $\mathrm{TE}_{011}$ mode, characterized by an azimuthal electric field structure, within the single-particle approximation. The temporal growth rate of the external magnetic field is adjusted to maintain electrons within the acceleration regime. The electron trajectory, energy, and phase shift between the transverse velocity and the electric field are obtained by numerically solving the relativistic Newton–Lorentz equation using a finite-difference scheme. This configuration enables efficient two-dimensional confinement and autoresonant acceleration of electrons, opening new perspectives for the development of compact and tunable high-energy electron sources.

\section{\label{sec:Formalism}THEORETICAL FORMALISM}

\subsection{\label{sec:Physchem}Physical scheme}
\begin{figure}[!htb]
	\centering
	\includegraphics*[width=.7\columnwidth]{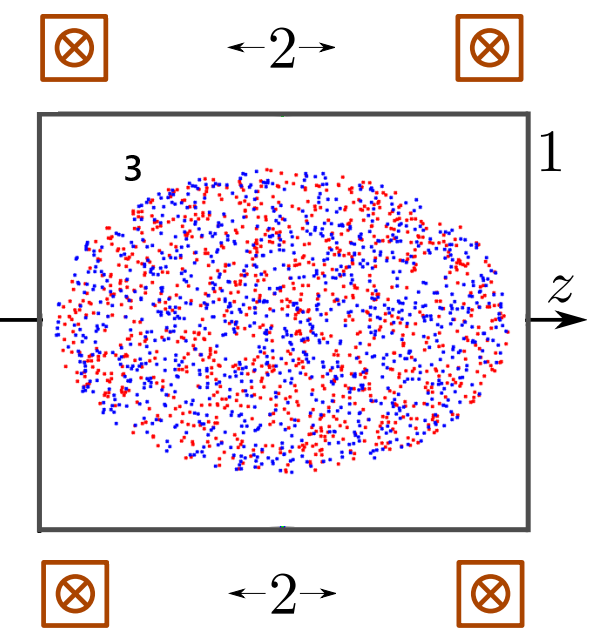}
	\caption{Simplified representation of the proposed configuration. The cylindrical cavity (1) is modeled as an ideal conductor and placed within a magnetic mirror trap formed by two direct-current coils (2). A confined hydrogen plasma (3) interacts with the microwave field through the GYRAC mechanism. Red and blue dots represent electrons and ions, respectively.}
	\label{fig:PhysicalScheme}
\end{figure}

To investigate the physical process of electron acceleration through the GYRAC mechanism based on the axisymmetric $\mathrm{TE}_{011}$ mode, we consider the conceptual setup illustrated in Fig.~\ref{fig:PhysicalScheme}. Although not shown in the figure, the excitation of this mode can be achieved using a suitable microwave coupling system. Various methods have been proposed in the literature depending on the specific application and cavity geometry \cite{joshi2019design, vennemann2015construction, guo2018using}.

\noindent In the present configuration, the cylindrical cavity (1) is modeled as an ideal conductor, enabling the use of analytical field expressions for the $\mathrm{TE}_{011}$ mode. While a real cavity might be constructed from 304 stainless steel and coated with silver to reduce wall losses, such material properties are not considered in this study. The cavity is placed within a mirror magnetic trap formed by two direct-current coils (2). A confined hydrogen plasma (3) is heated by the microwave field through the GYRAC mechanism, in which the ECR condition is sustained by gradually increasing the magnetic field intensity over time, controlled by adjusting the current in the coils (2).

\subsection{\label{sec:Theory}Theory and Simulation Model}
This work focuses on investigating the gyroresonant acceleration of electrons by a microwave field in the $\mathrm{TE}{011}$ mode, employing the single-particle approximation within a simplified model. In this model, the electron is treated as a single particle undergoing two-dimensional motion in the transverse mid-plane of the cavity ($z = L_c / 2$), where $L_c$ is the cavity length. Although this approach does not capture the full complexity of real plasma behavior, it effectively illustrates the fundamental mechanisms underlying autoresonant acceleration in the $\mathrm{TE}{011}$ mode.

\begin{figure}[!htb]
	\centering
	\includegraphics*[width=1.0\columnwidth]{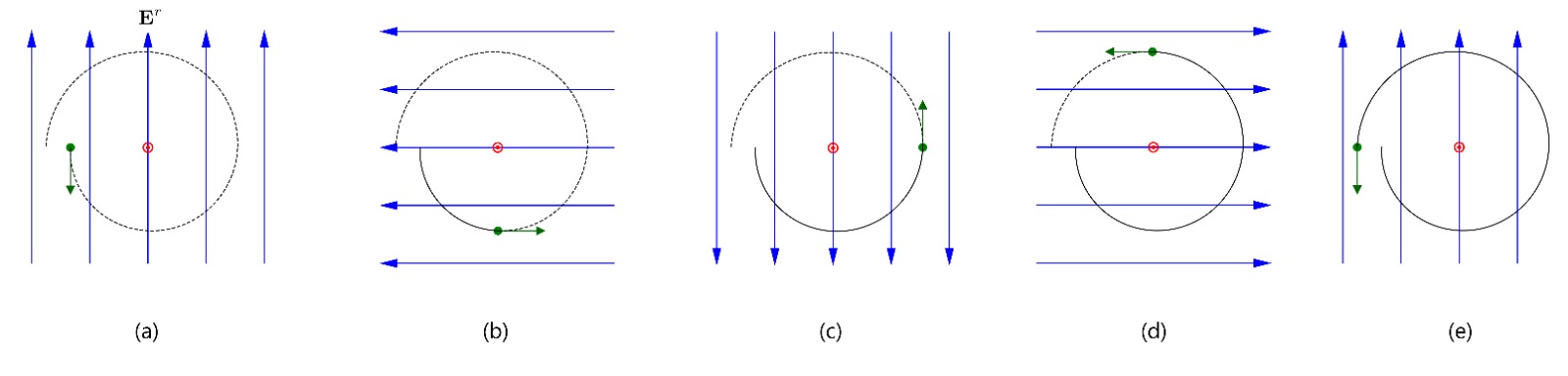}
	\caption{ Electron resonance interaction with the right-hand polarization (RHP) component at different time instants: (a) \( t=0 \), (b) \( t=T/4 \), (c) \( t=T/2 \), (d) \( t=3T/4 \), and (e) \( t=T \), where \( T = 2\pi / \omega \) represents the period of the electromagnetic wave. The blue arrows depict the right-handed electric field component of the high-frequency field, the green arrow indicates the electron velocity, and the red circle represents the magnetostatic field oriented out of the page. The dashed line illustrates the complete electron trajectory over one period of the electromagnetic wave, while the solid black line represents the electron’s path from \( t=0 \) to the indicated time instant. The Larmor radius increases as the electron gains energy.}
	\label{fig:ECR_Righ_Hand}
\end{figure}

This acceleration mechanism relies on the electron cyclotron resonance (ECR) phenomenon, which occurs when an electron in an external magnetic field efficiently absorbs energy from an electromagnetic wave with right-hand circular polarization (See Fig. \ref{fig:ECR_Righ_Hand}). 

In a uniform magnetic field, an electron undergoes circular motion at the cyclotron frequency given by  
\begin{equation}\label{eq:Omega}
\Omega_{c} = \frac{e B}{m_{e} \gamma},
\end{equation}  
where \( e \) is the electron charge, \( m_e \) its mass, \( B \) the magnetic field strength, and \( \gamma = (1 - \beta^2)^{-1/2} \) is the relativistic Lorentz factor. Here, \( \beta = v / c \) denotes the ratio of the electron's velocity \( v \) to the speed of light \( c \) in vacuum.\\

\noindent The phenomenon of autoresonance involves maintaining a continuous electron cyclotron resonance (ECR) interaction. When an electron interacts with a standing electromagnetic wave, the resonance condition \( \Omega_c = \omega \) can be sustained by gradually increasing the magnetic field strength to compensate for the rise in the Lorentz factor \( \gamma \). This compensation ensures that the electron remains in resonance with the wave, allowing it to continuously absorb energy. Such sustained interaction is essential in applications requiring high-energy electrons, as it enables consistent and efficient energy transfer.\\

\textbf{The GYRAC Mechanism and Phase Focusing}\\

In the scenario of temporal autoresonance, commonly referred to as GYRAC, a uniformly time-varying magnetic field is strategically employed. The electron cyclotron frequency adapts to the evolving magnetic field, which can be expressed as  
\begin{equation}
\mathbf{B} = B_0[1 + b(t)]\, \hat{\mathbf{k}},
\end{equation}
where \( B_0 = \omega m_e / e \) defines the classical resonance magnetic field, and \( b(t) \) is a dimensionless function that increases monotonically over time, reflecting the adjustments required to maintain resonance with the electromagnetic wave of frequency \( \omega \).

\begin{figure}[!htb]
	\centering
	\includegraphics*[width=0.75\columnwidth]{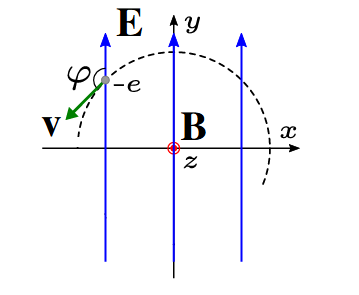}
	\caption{Spatial relationship between the particle trajectory, right-hand circularly polarized electric field $\mathbf{E}$, and external magnetic field B. The angle $\varphi$ represents the phase shift between the electron's velocity vector $\mathbf{v}$ and the electric field vector $\mathbf{E}$.}
	\label{fig:GYRAC_Resonant_Interaction}
\end{figure}

Electrons gain energy through their interaction with the right-hand circularly polarized electric field, expressed in cylindrical coordinates as  
\begin{equation}
\mathbf{E} = E_0 (\sin \varphi\, \hat{\mathbf{r}} + \cos \varphi\, \hat{\boldsymbol{\theta}}),
\end{equation}  
where \( \varphi \) represents the phase shift between the electric field vector and the electron's velocity vector. This interaction, which plays a key role in the energy transfer process, is illustrated in Fig.~\ref{fig:GYRAC_Resonant_Interaction}.

Considering the right-hand circularly polarized $\mathrm{TE}_{111}$ cylindrical mode, which can be approximated as a standing plane wave, Golovanivsky developed a model of two-dimensional relativistic dynamics \cite{golovanivsky1980autoresonant,golovanivsky1982gyrac, golovanivsky1983gyromagnetic}. This model led to the formulation of a set of differential equations that are fundamental for analyzing the evolution of energy and phase shift in charged particles. These equations, presented in Eqs.~(\ref{eq:GammaEvolution}) and (\ref{eq:VarphiEvolution}), provide key insights into the temporal behavior of particle dynamics:

\begin{equation}\label{eq:GammaEvolution}
\dot{\gamma} = -\,g_{0}\,\left(1\,-\,\frac{1}{\gamma^{2}}\right)^{1/2}\,\cos{\varphi},
\end{equation}

\begin{equation}\label{eq:VarphiEvolution}
\dot{\varphi} = \Bigl[b\left(\tau\right) -\left( \gamma - 1\right) \Bigr]\,\frac{1}{\gamma} + g_{0}\,\left(\gamma^{2}-1\right)^{-1/2}\,\sin{\varphi},
\end{equation}

where \( g_0 = \frac{E_0}{B_0 c} \) is the normalized electric field strength, and \( \tau = \omega t \) is the dimensionless time variable. According to Eq.~(\ref{eq:GammaEvolution}), the electron gains energy when \( \frac{\pi}{2} < \varphi < \frac{3\pi}{2} \), with the resonance condition corresponding to \( \varphi = \pi \). This result is essential for optimizing the electron acceleration within the GYRAC framework.

The GYRAC mechanism also exhibits a phase-focusing effect, wherein particles are consistently captured into the acceleration regime—commonly referred to as the GYRAC regime—regardless of their initial phase shift. This effect is particularly pronounced at low energies, where \( \gamma \approx 1 \).

From Eq.~(\ref{eq:VarphiEvolution}), effective trapping of particles within the GYRAC regime requires the magnetic field to increase according to the linear expression:  
\begin{equation}
b(\tau) = \alpha \tau,
\end{equation}  
where \( \alpha \) denotes the rate of change of the magnetic field and must satisfy the constraint:  
\begin{equation} \label{eq:alpha_cosntrain}
\alpha \leq 1.19\, g_0^{4/3}.
\end{equation}

This condition sets a critical upper bound on the allowable magnetic field growth rate to ensure efficient phase capture and sustained acceleration within the GYRAC regime. Enforcing this constraint is essential for maintaining the autoresonance condition across a broad range of particle energies and operational parameters.

In the present study, the cylindrical microwave mode $\mathrm{TE}_{011}$ is employed. This mode differs significantly from those that permit a stationary plane wave approximation, such as the $\mathrm{TE}_{111}$ mode. Due to its inherent spatial non-uniformities in both magnitude and direction of the electric field, the $\mathrm{TE}_{011}$ mode requires a localized approximation rather than a global one. These non-uniformities directly impact the conditions necessary for establishing and sustaining the GYRAC regime (see Eq.~\ref{eq:alpha_cosntrain}).

Unlike configurations in which a simple stationary plane wave model is applicable, the $\mathrm{TE}_{011}$ mode demands a more detailed analysis of the spatial field structure to accurately describe and control the dynamics of particles. The inhomogeneous nature of the electric field leads to a modified trapping condition, where conventional parameters defining the GYRAC regime must be adjusted. Such adjustments are essential because they directly influence the phase-focusing and autoresonance conditions that underlie efficient particle acceleration.

The electric and magnetic field components of the $\mathrm{TE}_{011}$ mode in a cylindrical cavity are expressed as follows:

\begin{equation}\label{eq:E_total} 
E_{\theta}^{\,hf}\left(r,\theta,z;t\right) = E_{0} \frac{J_{1}\left(k_{\perp}r\right)}{J_{1}(p_{01})} \sin{\left(k_{z}z\right)} \cos{\left(\omega t + \psi_0\right)},
\end{equation}

\begin{equation}\label{eq:B_r} 
B_{r}^{\,hf}\left(r,\theta,z;t\right) = E_{0} \frac{k_{z}}{\omega} \frac{J_{1}\left(k_{\perp}r\right)}{J_{1}(p_{01})} \cos{\left(k_{z}z\right)} \sin{\left(\omega t + \psi_0\right)},
\end{equation}

\begin{equation}\label{eq:B_z} 
B_{z}^{\,hf}\left(r,\theta,z;t\right) = -E_{0} \frac{k_{\perp}}{\omega} \frac{J_{0}\left(k_{\perp}r\right)}{J_{1}(p_{01})} \sin{\left(k_{z}z\right)} \sin{\left(\omega t + \psi_0\right)},
\end{equation}

where \( E_0 \) is the amplitude of the high-frequency electric field, and \( J_0 \) and \( J_1 \) denote the Bessel functions of the first kind of order zero and one, respectively. The radial wavenumber is given by \( k_{\perp} = q_{01} / R_c \), where \( R_c \) is the cavity radius and \( q_{01} = 3.83171 \) is the first zero of \( J_1(u) \). The axial wavenumber is \( k_z = \pi / L_c \), with \( L_c \) being the cavity length. The resonant frequency is defined as \( \omega = c \sqrt{k_{\perp}^2 + k_z^2} \), corresponding to the $\mathrm{TE}_{011}$ mode, and \( \psi_0 \) is an arbitrary phase.

The term \( J_1(p_{01}) \), where \( p_{01} = 1.84118 \), serves as a normalization factor ensuring that \( E_0 \) is properly scaled. The $\mathrm{TE}_{011}$ mode exhibits a high quality factor compared to other low-order cylindrical modes, enhancing its suitability for various applications \cite{pozar2021microwave}.

\begin{figure}[!htb]
\centering
\includegraphics[width=0.7\columnwidth]{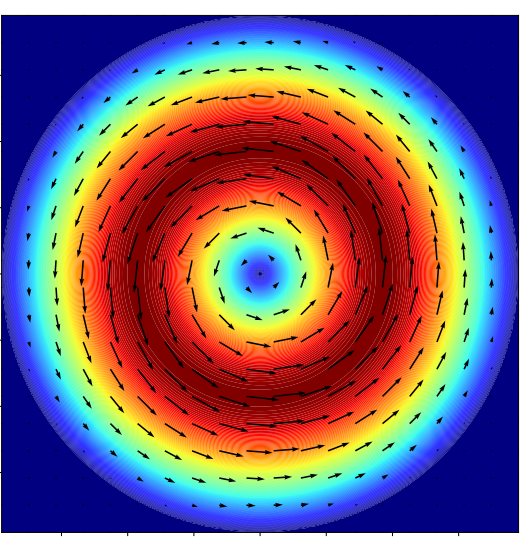}
\caption{Vector field and electric field magnitude of the $\mathrm{TE}_{011}$ mode at the plane \( z = L_c/2 \).}
\label{fig:TE011}
\end{figure}

\begin{figure}[!htb]
\centering
\includegraphics[width=1.0\columnwidth]{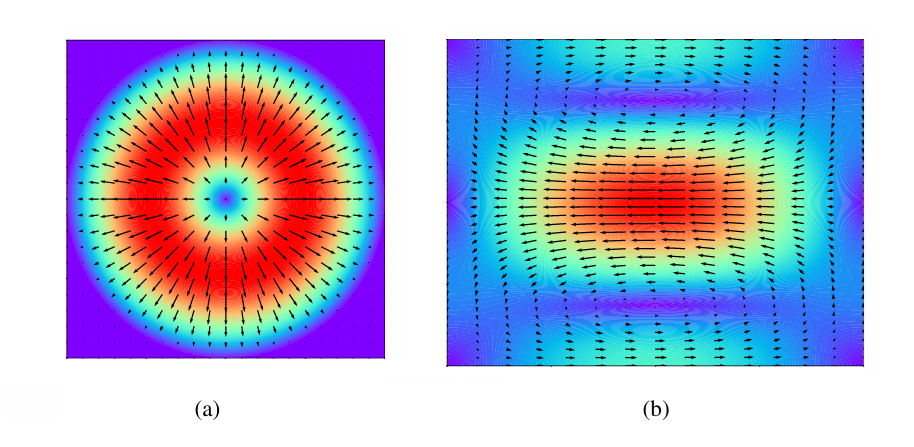}
\caption{Vector field and magnetic field magnitude of the $\mathrm{TE}_{011}$ mode: (a) at the plane \( z = L_c/2 \), and (b) at the plane \( x = 0 \).}
\label{fig:Bhf_TE011}
\end{figure}

The electric field distribution of the $\mathrm{TE}_{011}$ mode at the plane \( z = L_c/2 \) is shown in Fig.~\ref{fig:TE011}, while the magnetic field distribution at the planes \( z = L_c/2 \) and \( x = 0 \) is presented in Fig.~\ref{fig:Bhf_TE011}.

The microwave electric field can locally be described as the superposition of two circularly polarized components: one right-handed and one left-handed. The total high-frequency electric field is expressed as 
\begin{equation}
\mathbf{E}^{\mathrm{hf}} = \mathbf{E}^{l} + \mathbf{E}^{r},   
\end{equation}
where the left-handed and right-handed components are given, respectively, by:

\begin{equation} \label{E_left}
\mathbf{E}^{l} = \frac{E_{0}}{2} \frac{J_{1}\left(k_{\perp} r\right)}{J_{1}(p_{01})} \sin \left(k_z z\right) \left[\sin(\beta - \theta)\, \mathbf{\hat{i}} + \cos(\beta - \theta)\, \mathbf{\hat{j}} \right],
\end{equation}

\begin{equation} \label{E_right}
\mathbf{E}^{r} = \frac{E_{0}}{2} \frac{J_{1}\left(k_{\perp} r\right)}{J_{1}(p_{01})} \sin \left(k_z z\right) \left[-\sin(\beta + \theta)\, \mathbf{\hat{i}} + \cos(\beta + \theta)\, \mathbf{\hat{j}} \right],
\end{equation}

with \( \beta = \omega t + \psi_0 \), where \( \omega \) is the angular frequency and \( \psi_0 \) is an arbitrary phase.\\

\noindent \textbf{Refinement of the Constraint for Effective Trapping in the GYRAC Regime Enabled by the TE\(_{011}\) Mode}\\

To refine the constraint for effective trapping within the GYRAC regime driven by the $\mathrm{TE}_{011}$ mode, it is essential to account for the fact that the phase of the right-hand circularly polarized component of the electric field, \( \mathbf{E}^r \), varies not only in time but also in space. This spatial variation arises from the electron’s azimuthal drift, or guiding center motion, particularly along the \( \hat{\boldsymbol{\theta}} \) direction, as observed in the simulation results. This motion introduces additional complexity into the particle dynamics.

For an electron located at \( (r, \theta, z) \), which primarily interacts with the right-hand circularly polarized component \( \mathbf{E}^r \) of the microwave field, the effective trapping condition is locally adapted as:

\begin{equation} \label{eq:constrain_rz}
\alpha(r, z) \leq 1.19\, g_0^{4/3}(r, z),
\end{equation}

where the spatially dependent normalized electric field strength is given by:

\begin{equation} \label{eq:g0_rz}
g_0(r, z) = \frac{E_0}{2} \frac{J_1\left(k_{\perp} r\right)}{J_1\left(p_{01}\right)} \sin \left(k_z z\right).
\end{equation}

This refinement highlights the necessity of incorporating spatial variability into the model to ensure accurate prediction and control of the conditions required for effective phase trapping and sustained autoresonance within the GYRAC regime.\\

\noindent \textbf{Computational Analysis of Charged Particle Dynamics in the $\mathrm{TE}_{01p}$ Mode}\\

A comprehensive analysis of the dynamics of a charged particle that is autoresonantly accelerated by a microwave field in the $\mathrm{TE}_{01p}$ mode requires numerical techniques and computer-based simulations. The motion of the electron is governed by the relativistic Newton–Lorentz equation, which can be expressed in dimensionless form as follows \cite{birdsall2004plasma}:

\begin{equation}\label{eq:NLR}
\frac{d\mathbf{u}}{d\tau} = \mathbf{g}_{0} + \frac{\mathbf{u}}{\gamma} \times \mathbf{b},
\end{equation}

\noindent where \( \mathbf{u} = \mathbf{p} / mc \) is the normalized electron momentum, \( \mathbf{g}_0 = -\mathbf{E} / (B_0 c) \) is the normalized electric field, and \( \mathbf{b} = -\mathbf{B} / B_0 \) is the normalized total magnetic field, with \( \mathbf{B} = \mathbf{B}^{\,hf} + \mathbf{B}^{ext} \). The variable \( \tau = \omega t \) denotes the dimensionless time, and \( \gamma = \sqrt{1 + u^2} \) is the relativistic Lorentz factor.

For numerical integration, Eq.~(\ref{eq:NLR}) is discretized using a finite-difference scheme:

\begin{equation}\label{eq:NL_DF}
\frac{\mathbf{u}^{n+1/2} - \mathbf{u}^{n-1/2}}{\Delta \tau} = \mathbf{g}_0^n + \frac{\mathbf{u}^{n+1/2} + \mathbf{u}^{n-1/2}}{2 \gamma^n} \times \mathbf{b}^n,
\end{equation}

where \( n \) is the temporal step index. This equation is solved numerically using the Boris algorithm, a widely used method known for its accuracy, simplicity, and numerical stability in particle-in-cell (PIC) simulations \cite{qin2013boris}.

The particle's position is updated according to:

\begin{equation}
\mathbf{r}^{n+1} = \mathbf{r}^n + \frac{\mathbf{u}^{n+1/2}}{\gamma^{n+1/2}} \Delta \tau,
\end{equation}

where \( \gamma^{n+1/2} = \left[1 + (\mathbf{u}^{n+1/2})^2\right]^{1/2} \). This formulation ensures that relativistic effects are properly accounted for in the trajectory calculation. Such treatment is essential for maintaining the accuracy and physical consistency of simulations under varying electromagnetic field conditions.

\section{\label{sec:Results}RESULTS AND DISCUSSIONS}

The simulation of electron acceleration in the $\mathrm{TE}_{011}$ mode of a microwave field was performed using the following parameters:

\begin{itemize}
    \item \textbf{Cavity radius:} {7.84}{cm}
    \item \textbf{Cavity length:} {20.0}{cm}
    \item \textbf{Microwave frequency:} \(2.45\, \text{GHz}\)
    \item \textbf{Electric field strength:} \(1\, \text{kV/cm}\)
    \item \textbf{Time step (\(\Delta t\)):} \(0.8\, \text{ps}\)
    \item \textbf{Parameter \( \alpha \), controlling the temporal increase of the external magnetic field:} \(10^{-4} \leq \alpha \leq 3 \times 10^{-4}\)
\end{itemize}

To explore the dynamics under different spatial distributions, three distinct initial configurations were considered:

1. \textbf{Single electron.} A single electron, initially at rest, is released from a location at radial distance \(R_c/2\) on the plane \(z = L_c / 2\) (see Fig.~\ref{fig:TE011}). This case is designed to examine individual particle behavior under controlled and idealized conditions.

2. \textbf{Ring-like electron cloud.} A group of 1,000 electrons, all initially at rest, is uniformly distributed within the annular region defined by \(3R_c/8 < r < 9R_c/16\) on the plane \(z = L_c / 2\). This configuration enables the study of collective behavior without considering inter-electron interactions or self-consistent fields.

3. \textbf{Disk-like electron cloud.} Similarly, a set of 1,000 electrons, initially at rest, is distributed within the region \(r < R_c\) on the same plane \(z = L_c / 2\). This configuration represents a compact electron population and allows assessment of their overall response to the microwave field, again excluding inter-particle interactions and self-consistent fields.\\

\begin{figure}[!htb]
\centering
\includegraphics*[width=.75\columnwidth]{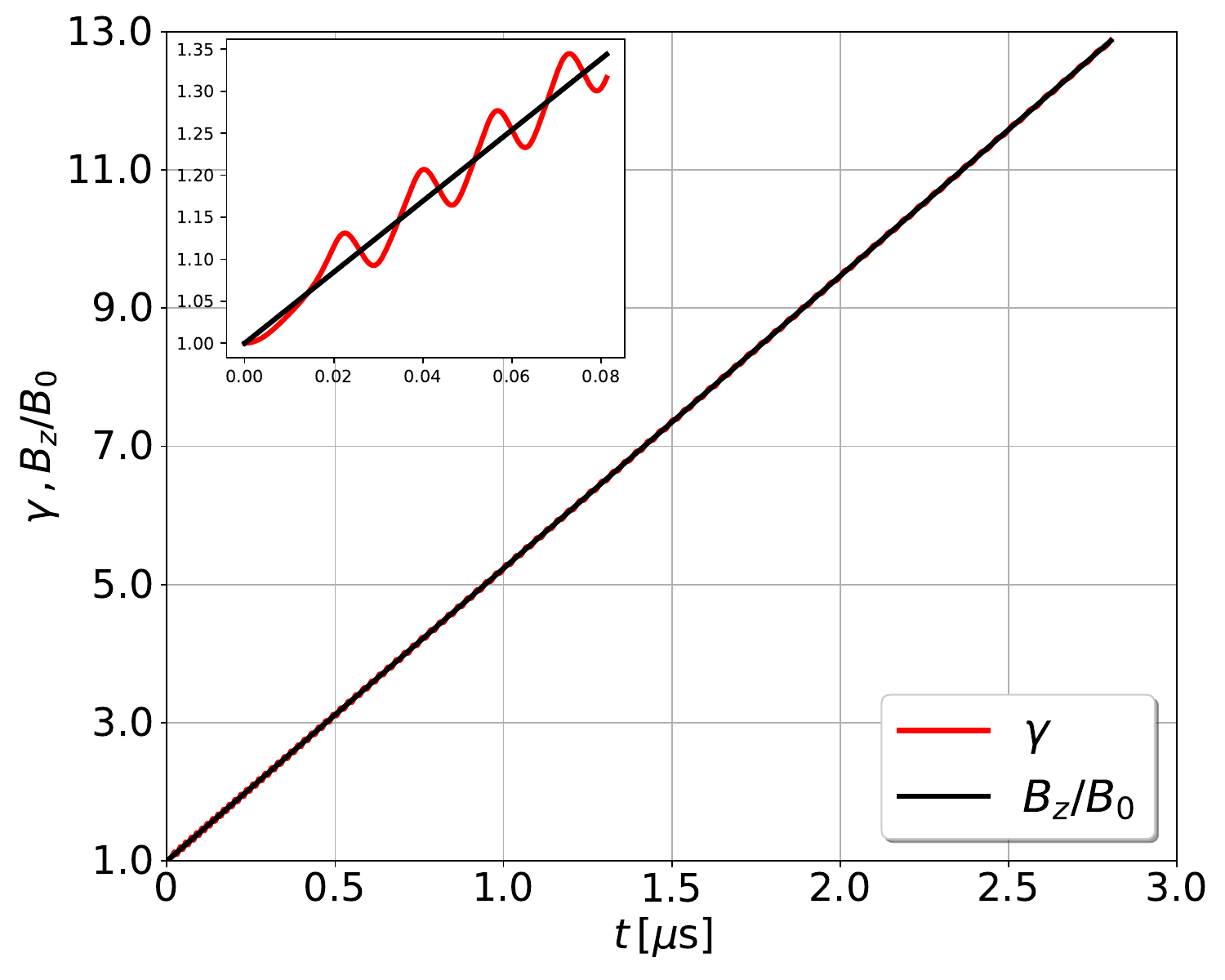}
\caption{Time evolution of the relativistic factor \( \gamma \) and the normalized magnetic field \( B_z / B_0 \) for \( \alpha = 2.75 \times 10^{-4} \). The inset provides a zoomed view highlighting the oscillations of \( \gamma \) around the linearly increasing magnetic field.}
\label{fig:Evolution_Gamma_B_T5}
\end{figure}

For the single-electron configuration, Fig.~\ref{fig:Evolution_Gamma_B_T5} shows the evolution of the relativistic factor \( \gamma \) and the external magnetic field up to the point where the electron impacts the cavity wall at \( t = {2.8}{\mu s} \). The red curve represents the increase in the electron's energy, while the black curve indicates the time-dependent growth of the external magnetic field that facilitates trapping within the GYRAC regime. Effective trapping is achieved because the phase shift \( \varphi \) remains predominantly within the \textit{acceleration band}, defined as the interval \( (\pi/2, 3\pi/2) \). 

\begin{figure}[!htb]
\centering
\includegraphics*[width=1.0\columnwidth]{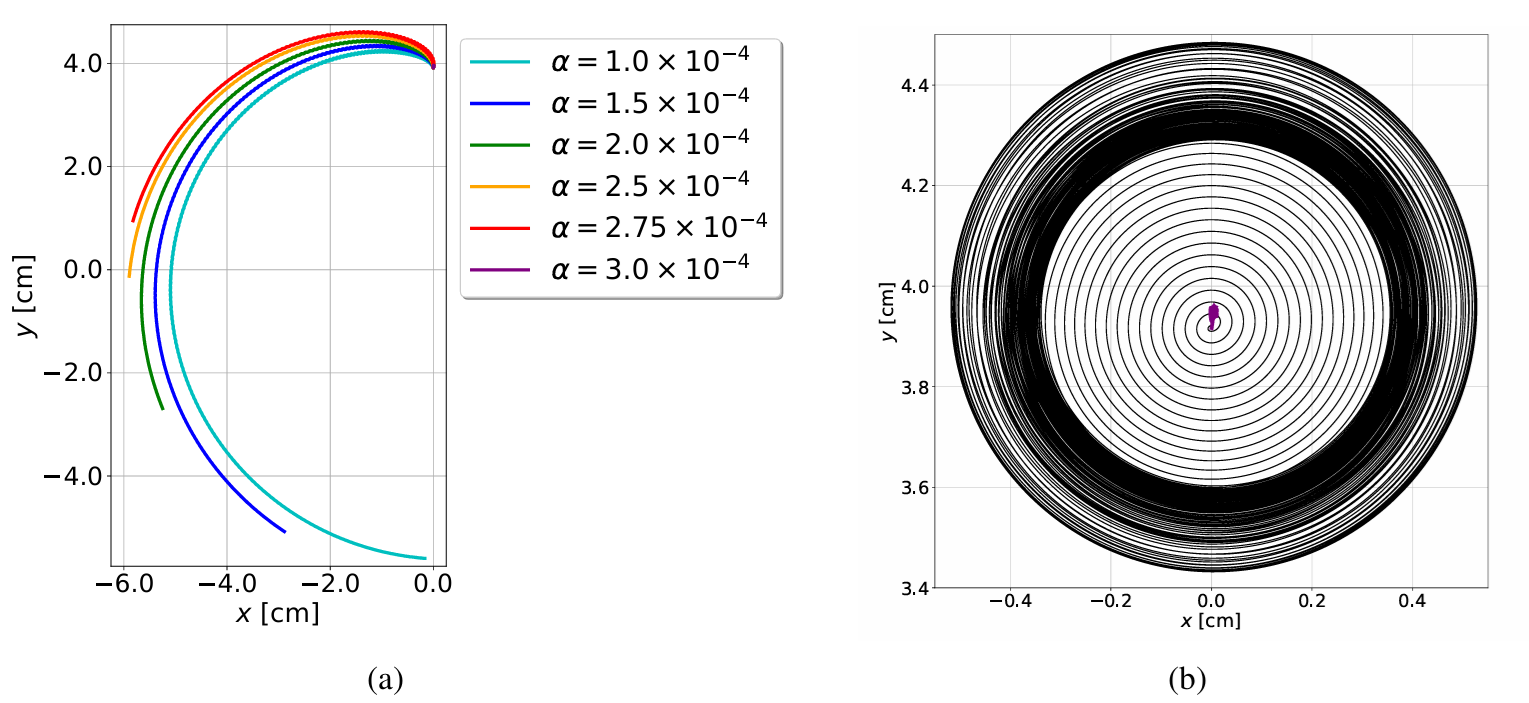}
\caption{(a) Trajectories of the guiding center illustrating the azimuthal drift motion of electrons for different values of the \( \alpha \) parameter, up to \( 4.65\, \mu\text{s} \). This panel highlights how variations in \( \alpha \) affect the displacement of the electron’s guiding center within the magnetic field. (b) Detailed view of the gyromotion of a single electron for \( \alpha = 3.0 \times 10^{-4} \), showing its rapid circular motion around the guiding center. For this value of \( \alpha \), the electron is not trapped in the GYRAC regime—as will be discussed later—and therefore does not exhibit drift motion.}
\label{fig:Electron_Drift}
\end{figure}

To elucidate the dynamics of electrons under varying electromagnetic conditions, Fig.~\ref{fig:Electron_Drift} provides a comprehensive visualization of particle motion influenced by the parameter \( \alpha \) within the GYRAC regime. The resulting trajectories exhibit two distinct components: the primary cyclotron motion around the guiding center and a drift motion in the \( \hat{\boldsymbol{\theta}} \) direction.

Panel (a) of Fig.~\ref{fig:Electron_Drift} displays the trajectories of the guiding centers for electrons subject to different values of \( \alpha \), highlighting the variation in azimuthal drift velocity. Notably, this drift becomes more pronounced for particles effectively trapped in the GYRAC.

Panel (b) focuses on the detailed gyromotion of a single electron for \( \alpha = 3.0 \times 10^{-4} \). This panel captures the rapid circular motion of the electron around its guiding center, offering an in-depth view of individual particle dynamics under controlled simulation conditions.

\begin{figure}[!htb]
\centering
\includegraphics*[width=0.8\columnwidth]{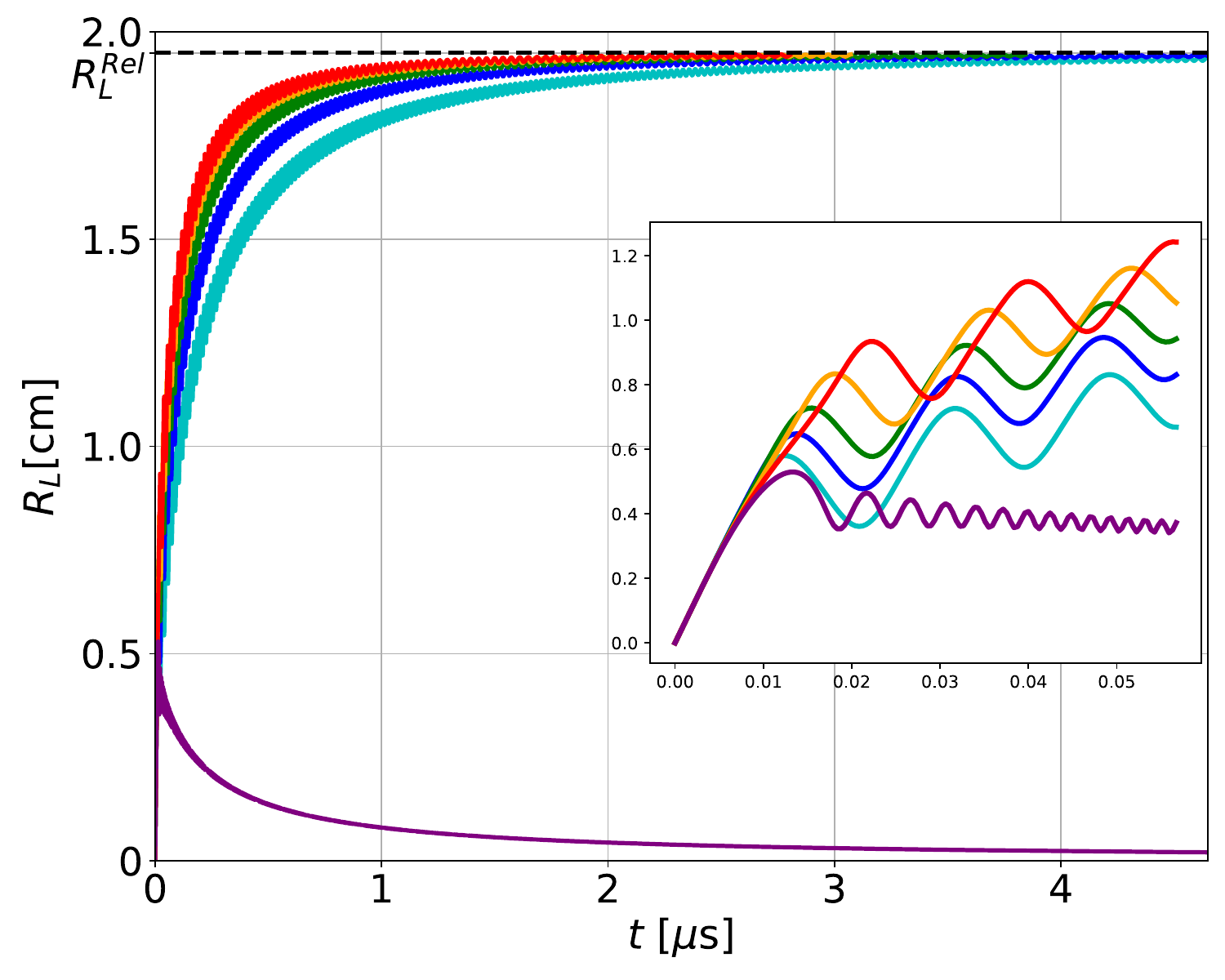}
\caption{Time evolution of the Larmor radius \( R_L \) for different values of the \( \alpha \) parameter. The graph shows how \( R_L \) increases as electrons gain energy at varying magnetic field ramp rates, eventually approaching the relativistic Larmor radius. Notably, the purple line corresponds to \( \alpha = 3.0 \times 10^{-4} \), and exhibits a decrease in \( R_L \) after \( 0.01\, \mu\text{s} \), indicating that in this case, the electron is not effectively trapped within the GYRAC regime.}

\label{fig:Evolution_RL_Ti}
\end{figure}

Figure~\ref{fig:Evolution_RL_Ti} presents the time evolution of the Larmor radius \( R_L \) for electrons under different values of the \( \alpha \) parameter. This visualization highlights the rapid increase of \( R_L \) across all cases, except for \( \alpha = 3.0 \times 10^{-4} \), represented by the purple line. In this particular case, the Larmor radius decreases after \( 0.01\, \mu\text{s} \), indicating that the electron is not effectively trapped within the GYRAC regime. 

In general, as electrons gain energy, their Larmor radius asymptotically approaches the relativistic Larmor radius \( R_L^{\text{Rel}} = c / \omega \), which is approximately {1.95}{cm} for the conditions considered.

The insights provided by Figs.~\ref{fig:Electron_Drift} and \ref{fig:Evolution_RL_Ti} are critical to understanding the role of \( \alpha \) in optimizing electron capture within the GYRAC regime. These results demonstrate that variations in \( \alpha \) significantly affect both trapping efficacy and particle dynamics, underscoring the importance of precise control over system parameters for effective acceleration.

\begin{figure}[!htb]
\centering
\includegraphics*[width=.75\columnwidth]{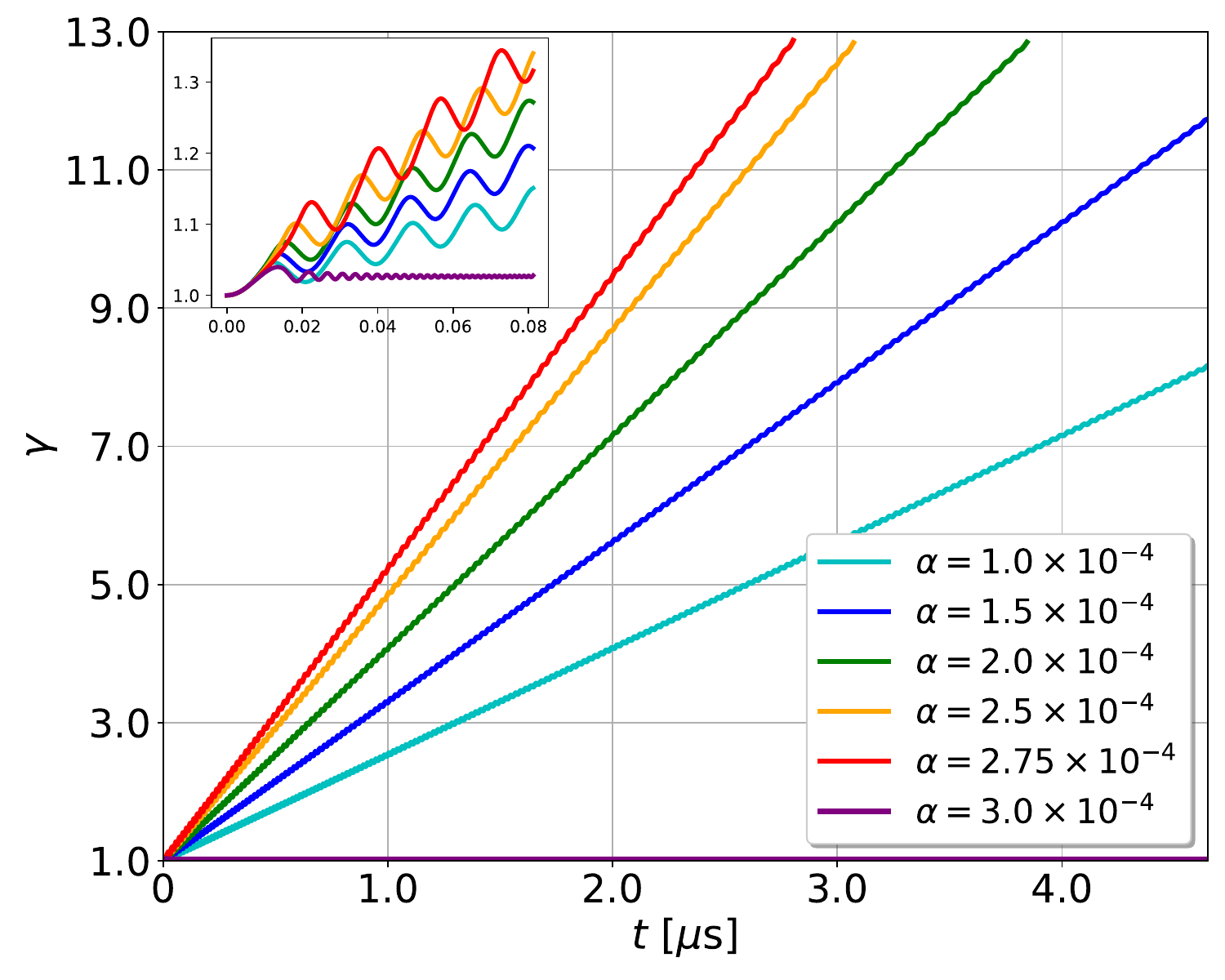}
\caption{Time evolution of the relativistic factor \( \gamma \) for different values of the \( \alpha \) parameter, illustrating how varying magnetic field ramp rates influence electron acceleration. Each curve corresponds to a different \( \alpha \) value and reflects the electron's energy gain under distinct trapping conditions.}
\label{fig:Evolution_Gamma_Ti}
\end{figure}

Figure~\ref{fig:Evolution_Gamma_Ti} presents the evolution of the electron energy for different values of the parameter \( \boldsymbol{\alpha} \), computed using the $g_0$ value given by Eq.~(\ref{eq:g0_rz}) at the position \( (r, z) = (R_c/2, L_c/2) \). All selected \( \alpha \) values satisfy the trapping condition defined in Eq.~(\ref{eq:constrain_rz}), except for \( \alpha = 3.0 \times 10^{-4} \), indicated by the purple line.

In this study, the trapping condition evolves over time due to two primary factors: spatial variations in the electric field strength, given by \( g_0 = g_0(r, z) \), and temporal shifts in the field phase caused by the electron's azimuthal drift motion (see Fig.~\ref{fig:Electron_Drift}a). Despite these complexities, the electron remains effectively trapped in the GYRAC regime for all considered cases (see Fig.~\ref{fig:Evolution_Gamma_Ti}).\\

\textbf{Results for Disk-Like Electron Cloud:}\\

\begin{figure}[!htb]
\centering
\includegraphics*[width=1.0\columnwidth]{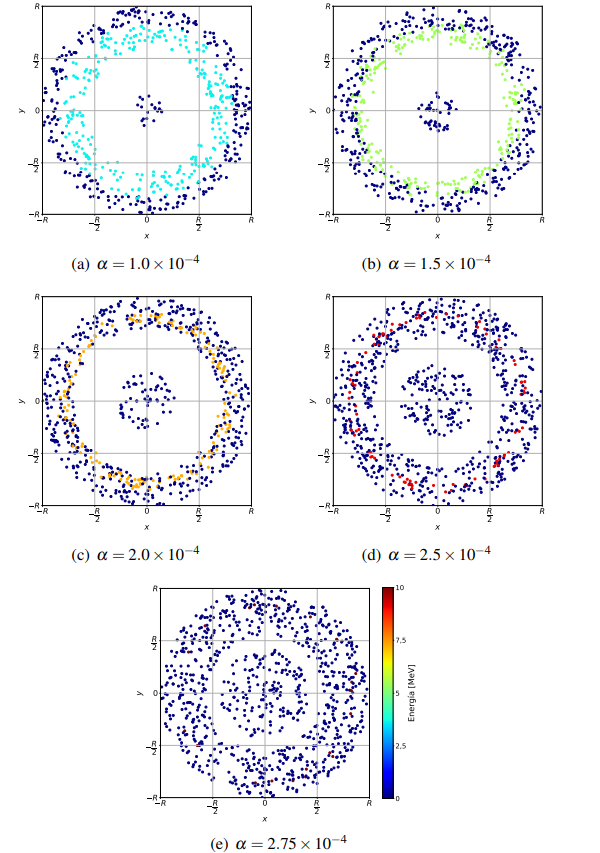}
\caption{Energy distribution of the disk-like electron cloud at \( t = 4.65\, \mu\text{s} \) for different values of the \( \alpha \) parameter, highlighting distinct groups of electrons based on their final energy levels.}
\label{fig:Disk_Evolution}
\end{figure}

In Fig.~\ref{fig:Disk_Evolution}, we examine the energy distribution of the disk-like electron cloud at \( t = 4.65\, \mu\text{s} \), for various values of the \( \alpha \) parameter. This visualization reveals that not all electrons are effectively trapped within the GYRAC regime, as different values of \( \alpha \) lead to varying trapping efficiencies.

The spatial non-uniformity of the electric field associated with the $\mathrm{TE}_{011}$ microwave mode plays a critical role in this variability, strongly influencing the trapping condition described by Eq.~(\ref{eq:constrain_rz}).

Based on their dynamic response, the electrons can be categorized into three distinct groups:

1. \textbf{Uncaptured Particles (UCP):} These electrons are not trapped in the autoresonant regime, primarily due to a persistent phase mismatch. Their trajectories deviate from those required for sustained acceleration, resulting in their classification as uncaptured particles.

2. \textbf{Captured Particles (CP):} This group comprises electrons that reach high energies under favorable phase and field conditions. Their energy peaks reflect successful interaction with the GYRAC regime, indicating effective trapping and acceleration.

3. \textbf{Escaped Particles (EP):} These electrons impact the cavity wall, either because they were initially located too close to the boundary or because their Larmor radius increased rapidly—often due to drift motion. Such impacts typically lead to energy loss and removal from the acceleration process.

The data in Fig.~\ref{fig:Disk_Evolution} underscores the strong influence of the \( \alpha \) parameter on electron behavior and energy distribution. These results highlight the importance of precise electromagnetic field control to optimize trapping and acceleration efficiency within the GYRAC regime.

The outcomes of five numerical experiments, each performed with a different value of the \( \alpha \) parameter, are summarized in Table I. For each case, three independent simulations were conducted. The table reports the corresponding fractions of captured particles (CP), uncaptured particles (UCP), and escaped particles (EP)—those that impacted the cavity wall—evaluated at \( t = 4.65\, \mu\text{s} \).\\

\begin{table}[h!] \label{Table:Disk-Cloud}
\setlength\tabcolsep{3.5pt}
\centering
\caption{\textbf{Outcomes for Disk-Like Electron Clouds Under Different \(\alpha\) Parameters.} The table shows the fractions of captured particles (CP), uncaptured particles (UCP), and escaped particles (EP) for various values of the magnetic field ramp rate \( \alpha \). Each case was evaluated at \( t = 4.65\, \mu\text{s} \), and the values correspond to averages over three independent simulations.}
\begin{tabular}{|c|c|c|c|c|}
\hline$\alpha$ & Exp. & $\%$ \textbf{CP} & $\%$ \textbf{UCP} & $\%$ \textbf{EP} \\
\hline \multirow{3}{*}{$1.0 \times 10^{-4}$} & 1 & 29.0 & 32.3 & 38.7 \\
\cline { 2 - 5 } & 2 & 28.6 & 34.0 & 37.4 \\
\cline { 2 - 5 } & 3 & 29.4 & 34.9 & 35.7 \\
\hline \multirow{3}{*}{$1.5 \times 10^{-4}$} & 1 & 26.0 & 43.7 & 30.3 \\
\cline { 2 - 5 } & 2 & 21.5 & 44.8 & 33.7 \\
\cline { 2 - 5 } & 3 & 22.9 & 44.7 & 32.4 \\
\hline \multirow{3}{*}{$2.0 \times 10^{-4}$} & 1 & 16.7 & 56.5 & 26.8 \\
\cline { 2 - 5 } & 2 & 17.9 & 57.2 & 24.9 \\
\cline { 2 - 5 } & 3 & 15.0 & 56.1 & 28.9 \\
\hline \multirow{3}{*}{$2.5 \times 10^{-4}$} & 1 & 8.9 & 73.5 & 17.6 \\
\cline { 2 - 5 } & 2 & 8.7 & 74.1 & 17.2 \\
\cline { 2 - 5 } & 3 & 7.7 & 74.2 & 18.1 \\
\hline & 1 & 2.1 & 86.2 & 11.7 \\
\cline { 2 - 5 } $2.75 \times 10^{-4}$ & 2 & 1.8 & 87.8 & 10.4 \\
\cline { 2 - 5 } & 3 & 1.9 & 87.5 & 10.6 \\
\hline
\end{tabular}
\end{table}

\textbf{Results for Ring-Like Electron Cloud:} \\

\begin{figure}[!htb]
\centering
\includegraphics*[width=1.0\columnwidth]{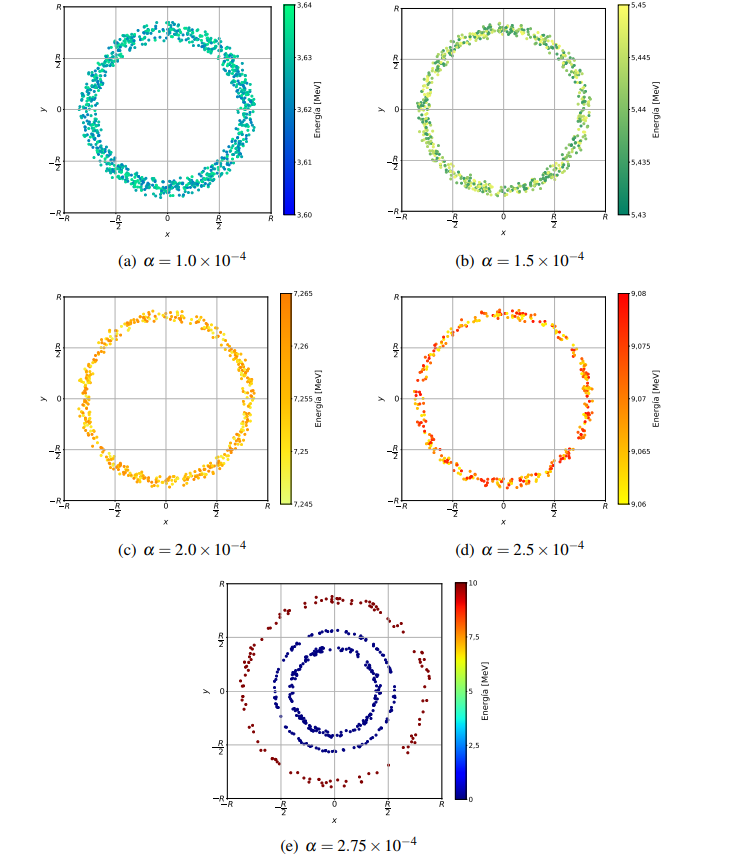}
\caption{
Energy distribution of the ring-like electron cloud at \( t = 4.65\, \mu\text{s} \) for different values of the \( \alpha \) parameter. The graph illustrates how variations in \( \alpha \) affect trapping efficiency within the GYRAC regime, as reflected by the grouping of electrons according to their final energy levels.
}
\label{fig:Ring_Evolution}
\end{figure}

Figure~\ref{fig:Ring_Evolution} presents the energy distribution of the ring-like electron cloud after \( 4.65\, \mu\text{s} \) of interaction with the microwave field under various values of the \( \alpha \) parameter. 

This figure shows how variations in the magnetic field ramp rate ($\alpha$) influence the energy distribution of the electron cloud, thereby directly affecting trapping efficiency within the GYRAC regime.

The plot categorizes electrons into distinct groups based on their final energy levels, revealing a clear correlation between the value of \( \alpha \) and the resulting energy distribution. This classification allows for the identification of capture efficiency and highlights the conditions under which electrons remain confined within the GYRAC regime or escape due to insufficient trapping.

In this configuration, where the electron cloud is strategically positioned in a ring-like region, a high percentage of electrons are captured within the GYRAC regime (see Table II). This table reports the values of \( \alpha \) along with the corresponding fractions of captured particles (CP), uncaptured particles (UCP), and escaped particles (EP)—defined as those that impacted the cavity wall—evaluated at \( t = 4.65\, \mu\text{s} \).

\begin{table}[h!] \label{Table:Ring-Cloud}
\setlength\tabcolsep{3.5pt}
\centering
\caption{\textbf{Outcomes for Ring-Like Electron Clouds Under Different \(\alpha\) Parameters.} This table reports the fractions of captured particles (CP), uncaptured particles (UCP), and escaped particles (EP)—those that impacted the cavity wall—at \( t = 4.65\, \mu\text{s} \) for various values of the \( \alpha \) parameter. Each \( \alpha \) value was tested in three independent numerical experiments.}

\begin{tabular}{|c|c|c|c|c|}
\hline$\alpha$ & Exp. & $\%$ PC & $\%$ PSC & $\%$ PE \\
\hline \multirow{3}{*}{$1.0 \times 10^{-4}$} & 1 & 95.9 & 0 & 4.1 \\
\cline { 2 - 5 } & 2 & 95.4 & 0 & 4.6 \\
\cline { 2 - 5 } & 3 & 95.4 & 0 & 4.6 \\
\hline \multirow{3}{*}{$1.5 \times 10^{-4}$} & 1 & 68.4 & 0 & 31.6 \\
\cline { 2 - 5 } & 2 & 69.2 & 0 & 30.8 \\
\cline { 2 - 5 } & 3 & 68.7 & 0 & 31.3 \\
\hline \multirow{3}{*}{$2.0 \times 10^{-4}$} & 1 & 48.8 & 0 & 51.2 \\
\cline { 2 - 5 } & 2 & 50.1 & 0 & 49.9 \\
\cline { 2 - 5 } & 3 & 49.0 & 0 & 51.0 \\
\hline \multirow{3}{*}{$2.5 \times 10^{-4}$} & 1 & 35.3 & 0 & 64.7 \\
\cline { 2 - 5 } & 2 & 37.0 & 0 & 63.0 \\
\cline { 2 - 5 } & 3 & 33.6 & 0 & 66.4 \\
\hline \multirow{3}{*}{$2.75 \times 10^{-4}$} & 1 & 12.3 & 29.2 & 58.5 \\
\cline { 2 - 5 } & 2 & 11.4 & 28.0 & 60.6 \\
\cline { 2 - 5 } & 3 & 10.4 & 34.2 & 55.4 \\
\hline
\end{tabular}
\end{table}

\section{\label{sec:conclusions}CONCLUSIONS}

The study confirms the critical role of the right-handed (RHP) component of a circularly polarized electric field in transferring energy to electrons, thereby enhancing the understanding of the physical mechanisms underlying autoresonant acceleration. Numerical experiments demonstrated that the GYRAC regime can be effectively established and sustained by appropriately tuning the growth rate of the external magnetic field, using the cylindrical $\mathrm{TE}_{011}$ mode of a microwave field.\\

It was shown that electrons can be accelerated under electron cyclotron resonance (ECR) conditions in time-varying magnetic fields using a cylindrical \( \mathrm{TE}_{011} \) mode, provided that resonance is maintained through careful control of key parameters. The energy levels reached in these simulations suggest promising applications of the GYRAC–\( \mathrm{TE}_{011} \) mechanism in the development of advanced radiation sources.\\

Additionally, a ring-shaped region was identified in which electrons are successfully captured within the self-resonance regime. The size and location of this region were found to depend sensitively on the value of the parameter \( \alpha \), which governs the time-dependent increase of the magnetic field.\\

Future work will focus on self-consistent simulations of hydrogen plasma confined in magnetic traps to further investigate GYRAC–\( \mathrm{TE}_{011} \) acceleration. These studies aim to establish a solid theoretical and computational foundation for the design of compact radiation sources based on this mechanism.\\

\textbf{CRediT authorship contribution statement}\\

\noindent \textbf{Oswaldo Otero:} Conceptualization, Formal analysis, Methodology.  \textbf{Jesús E. López:} Software, Validation. \textbf{P. Tsygankov:}   Methodology, Formal analysis, Writing- review \& editing.  \textbf{Carlos J. Páez-González:} Conceptualization, Formal analysis. \textbf{E. A. Orozco:} Conceptualization, Methodology, Formal analysis, Writing - original draft.\\\\

\author{Oswaldo Otero}
\email{oswaldoterolarte@gmail.com}
\author{Jesús E. López}
\author{P. Tsygankov}%
\author{Carlos J. Páez-González}%
\author{E. A. Orozco}%
\email{eaorozco@uis.edu.co}

\begin{acknowledgments}
This work was carried out with the support of the Departamento Administrativo de Ciencia,
Tecnología e Innovación, Colombia, and Ministerio de Ciencia Tecnología e Innovación,
Colombia, through the announcement No 852-2019 (1102-852-71985) and the Universidad Industrial de Santander (UIS), Colombia, (Project ID: 9482-2665).\\
\end{acknowledgments}

\appendix

\nocite{*}

\bibliography{apssamp}

\providecommand{\noopsort}[1]{}\providecommand{\singleletter}[1]{#1}%
\begin{thebibliography}{30}%
\makeatletter
\providecommand \@ifxundefined [1]{%
 \@ifx{#1\undefined}
}%
\providecommand \@ifnum [1]{%
 \ifnum #1\expandafter \@firstoftwo
 \else \expandafter \@secondoftwo
 \fi
}%
\providecommand \@ifx [1]{%
 \ifx #1\expandafter \@firstoftwo
 \else \expandafter \@secondoftwo
 \fi
}%
\providecommand \natexlab [1]{#1}%
\providecommand \enquote  [1]{``#1''}%
\providecommand \bibnamefont  [1]{#1}%
\providecommand \bibfnamefont [1]{#1}%
\providecommand \citenamefont [1]{#1}%
\providecommand \href@noop [0]{\@secondoftwo}%
\providecommand \href [0]{\begingroup \@sanitize@url \@href}%
\providecommand \@href[1]{\@@startlink{#1}\@@href}%
\providecommand \@@href[1]{\endgroup#1\@@endlink}%
\providecommand \@sanitize@url [0]{\catcode `\\12\catcode `\$12\catcode
  `\&12\catcode `\#12\catcode `\^12\catcode `\_12\catcode `\%12\relax}%
\providecommand \@@startlink[1]{}%
\providecommand \@@endlink[0]{}%
\providecommand \url  [0]{\begingroup\@sanitize@url \@url }%
\providecommand \@url [1]{\endgroup\@href {#1}{\urlprefix }}%
\providecommand \urlprefix  [0]{URL }%
\providecommand \Eprint [0]{\href }%
\providecommand \doibase [0]{https://doi.org/}%
\providecommand \selectlanguage [0]{\@gobble}%
\providecommand \bibinfo  [0]{\@secondoftwo}%
\providecommand \bibfield  [0]{\@secondoftwo}%
\providecommand \translation [1]{[#1]}%
\providecommand \BibitemOpen [0]{}%
\providecommand \bibitemStop [0]{}%
\providecommand \bibitemNoStop [0]{.\EOS\space}%
\providecommand \EOS [0]{\spacefactor3000\relax}%
\providecommand \BibitemShut  [1]{\csname bibitem#1\endcsname}%
\let\auto@bib@innerbib\@empty
\bibitem [{\citenamefont {Kolomenskii}\ and\ \citenamefont
  {A.N.Levedek}(1963)}]{Kolomenskii1963}%
  \BibitemOpen
  \bibfield  {author} {\bibinfo {author} {\bibfnamefont {A.}~\bibnamefont
  {Kolomenskii}}\ and\ \bibinfo {author} {\bibnamefont {A.N.Levedek}},\
  }\bibfield  {title} {\bibinfo {title} {Resonance effects associated with
  particle motion in a plane electromagnetic wave},\ }\href@noop {} {\bibfield
  {journal} {\bibinfo  {journal} {Soviet Physics JETP}\ }\textbf {\bibinfo
  {volume} {17}} (\bibinfo {year} {1963})}\BibitemShut {NoStop}%
\bibitem [{\citenamefont {Davydovski}(1962)}]{davydovski1962possibility}%
  \BibitemOpen
  \bibfield  {author} {\bibinfo {author} {\bibfnamefont {V.~Y.}\ \bibnamefont
  {Davydovski}},\ }\bibfield  {title} {\bibinfo {title} {On the possibility of
  accelerating charged particles by electromagnetic waves in a constant
  magnetic field},\ }\href@noop {} {\bibfield  {journal} {\bibinfo  {journal}
  {Zh. Eksperim. i Teor. Fiz.}\ }\textbf {\bibinfo {volume} {43}} (\bibinfo
  {year} {1962})}\BibitemShut {NoStop}%
\bibitem [{\citenamefont {Jory}\ and\ \citenamefont
  {Trivelpiece}(1968)}]{jory1968charged}%
  \BibitemOpen
  \bibfield  {author} {\bibinfo {author} {\bibfnamefont {H.}~\bibnamefont
  {Jory}}\ and\ \bibinfo {author} {\bibfnamefont {A.}~\bibnamefont
  {Trivelpiece}},\ }\bibfield  {title} {\bibinfo {title} {Charged-particle
  motion in large-amplitude electromagnetic fields},\ }\href@noop {} {\bibfield
   {journal} {\bibinfo  {journal} {Journal of Applied Physics}\ }\textbf
  {\bibinfo {volume} {39}},\ \bibinfo {pages} {3053} (\bibinfo {year}
  {1968})}\BibitemShut {NoStop}%
\bibitem [{\citenamefont {Milant'ev}(2013)}]{milant2013cyclotron}%
  \BibitemOpen
  \bibfield  {author} {\bibinfo {author} {\bibfnamefont {V.~P.}\ \bibnamefont
  {Milant'ev}},\ }\bibfield  {title} {\bibinfo {title} {Cyclotron
  autoresonance—50 years since its discovery},\ }\href@noop {} {\bibfield
  {journal} {\bibinfo  {journal} {Physics-Uspekhi}\ }\textbf {\bibinfo {volume}
  {56}},\ \bibinfo {pages} {823} (\bibinfo {year} {2013})}\BibitemShut
  {NoStop}%
\bibitem [{\citenamefont {Loeb}\ and\ \citenamefont
  {Friedland}(1986)}]{loeb1986autoresonance}%
  \BibitemOpen
  \bibfield  {author} {\bibinfo {author} {\bibfnamefont {A.}~\bibnamefont
  {Loeb}}\ and\ \bibinfo {author} {\bibfnamefont {L.}~\bibnamefont
  {Friedland}},\ }\bibfield  {title} {\bibinfo {title} {Autoresonance laser
  accelerator},\ }\href@noop {} {\bibfield  {journal} {\bibinfo  {journal}
  {Physical Review A}\ }\textbf {\bibinfo {volume} {33}},\ \bibinfo {pages}
  {1828} (\bibinfo {year} {1986})}\BibitemShut {NoStop}%
\bibitem [{\citenamefont {Salamin}\ \emph {et~al.}(2015)\citenamefont
  {Salamin}, \citenamefont {Li}, \citenamefont {Galow},\ and\ \citenamefont
  {Keitel}}]{salamin2015feasibility}%
  \BibitemOpen
  \bibfield  {author} {\bibinfo {author} {\bibfnamefont {Y.~I.}\ \bibnamefont
  {Salamin}}, \bibinfo {author} {\bibfnamefont {J.-X.}\ \bibnamefont {Li}},
  \bibinfo {author} {\bibfnamefont {B.~J.}\ \bibnamefont {Galow}},\ and\
  \bibinfo {author} {\bibfnamefont {C.~H.}\ \bibnamefont {Keitel}},\ }\bibfield
   {title} {\bibinfo {title} {Feasibility of electron cyclotron autoresonance
  acceleration by a short terahertz pulse},\ }\href@noop {} {\bibfield
  {journal} {\bibinfo  {journal} {Optics Express}\ }\textbf {\bibinfo {volume}
  {23}},\ \bibinfo {pages} {17560} (\bibinfo {year} {2015})}\BibitemShut
  {NoStop}%
\bibitem [{\citenamefont {Shpitalnik}\ \emph {et~al.}(1991)\citenamefont
  {Shpitalnik}, \citenamefont {Cohen}, \citenamefont {Dothan},\ and\
  \citenamefont {Friedland}}]{shpitalnik1991autoresonance}%
  \BibitemOpen
  \bibfield  {author} {\bibinfo {author} {\bibfnamefont {R.}~\bibnamefont
  {Shpitalnik}}, \bibinfo {author} {\bibfnamefont {C.}~\bibnamefont {Cohen}},
  \bibinfo {author} {\bibfnamefont {F.}~\bibnamefont {Dothan}},\ and\ \bibinfo
  {author} {\bibfnamefont {L.}~\bibnamefont {Friedland}},\ }\bibfield  {title}
  {\bibinfo {title} {Autoresonance microwave accelerator},\ }\href@noop {}
  {\bibfield  {journal} {\bibinfo  {journal} {Journal of applied physics}\
  }\textbf {\bibinfo {volume} {70}},\ \bibinfo {pages} {1101} (\bibinfo {year}
  {1991})}\BibitemShut {NoStop}%
\bibitem [{\citenamefont {McDermott}\ \emph {et~al.}(1985)\citenamefont
  {McDermott}, \citenamefont {Furuno},\ and\ \citenamefont
  {Luhmann~Jr}}]{mcdermott1985production}%
  \BibitemOpen
  \bibfield  {author} {\bibinfo {author} {\bibfnamefont {D.}~\bibnamefont
  {McDermott}}, \bibinfo {author} {\bibfnamefont {D.}~\bibnamefont {Furuno}},\
  and\ \bibinfo {author} {\bibfnamefont {N.}~\bibnamefont {Luhmann~Jr}},\
  }\bibfield  {title} {\bibinfo {title} {Production of relativistic, rotating
  electron beams by gyroresonant rf acceleration in a te111 cavity},\
  }\href@noop {} {\bibfield  {journal} {\bibinfo  {journal} {Journal of applied
  physics}\ }\textbf {\bibinfo {volume} {58}},\ \bibinfo {pages} {4501}
  (\bibinfo {year} {1985})}\BibitemShut {NoStop}%
\bibitem [{\citenamefont {Neishtadt}\ and\ \citenamefont
  {Timofeev}(1987)}]{neishtadt1987autoresonance}%
  \BibitemOpen
  \bibfield  {author} {\bibinfo {author} {\bibfnamefont {A.}~\bibnamefont
  {Neishtadt}}\ and\ \bibinfo {author} {\bibfnamefont {A.}~\bibnamefont
  {Timofeev}},\ }\bibfield  {title} {\bibinfo {title} {Autoresonance in
  electron cyclotron heating of a plasma},\ }\href@noop {} {\bibfield
  {journal} {\bibinfo  {journal} {Zh. Eksp. Teor. Fiz}\ }\textbf {\bibinfo
  {volume} {93}},\ \bibinfo {pages} {1706} (\bibinfo {year}
  {1987})}\BibitemShut {NoStop}%
\bibitem [{\citenamefont {Dugar-Zhabon}\ and\ \citenamefont
  {Orozco}(2009)}]{dugar2009cyclotron}%
  \BibitemOpen
  \bibfield  {author} {\bibinfo {author} {\bibfnamefont {V.~D.}\ \bibnamefont
  {Dugar-Zhabon}}\ and\ \bibinfo {author} {\bibfnamefont {E.~A.}\ \bibnamefont
  {Orozco}},\ }\bibfield  {title} {\bibinfo {title} {Cyclotron spatial
  autoresonance acceleration model},\ }\href@noop {} {\bibfield  {journal}
  {\bibinfo  {journal} {Physical Review Special Topics-Accelerators and Beams}\
  }\textbf {\bibinfo {volume} {12}},\ \bibinfo {pages} {041301} (\bibinfo
  {year} {2009})}\BibitemShut {NoStop}%
\bibitem [{\citenamefont {Dugar-Zhabon}\ and\ \citenamefont
  {Orozco}(2017)}]{dugar2017compact}%
  \BibitemOpen
  \bibfield  {author} {\bibinfo {author} {\bibfnamefont {V.}~\bibnamefont
  {Dugar-Zhabon}}\ and\ \bibinfo {author} {\bibfnamefont {E.}~\bibnamefont
  {Orozco}},\ }\href@noop {} {\bibinfo {title} {Compact self-resonant x-ray
  source, patent us9666403b2}} (\bibinfo {year} {2017})\BibitemShut {NoStop}%
\bibitem [{\citenamefont {Otero}\ and\ \citenamefont
  {Orozco}(2019)}]{otero2019numerical}%
  \BibitemOpen
  \bibfield  {author} {\bibinfo {author} {\bibfnamefont {O.}~\bibnamefont
  {Otero}}\ and\ \bibinfo {author} {\bibfnamefont {E.}~\bibnamefont {Orozco}},\
  }\bibfield  {title} {\bibinfo {title} {Numerical simulation of electron
  cyclotron resonance phenomenon using an axisymmetric transverse electric
  field},\ }in\ \href@noop {} {\emph {\bibinfo {booktitle} {Journal of Physics:
  Conference Series}}},\ Vol.\ \bibinfo {volume} {1386}\ (\bibinfo
  {organization} {IOP Publishing},\ \bibinfo {year} {2019})\ p.\ \bibinfo
  {pages} {012123}\BibitemShut {NoStop}%
\bibitem [{\citenamefont {Velazco}\ and\ \citenamefont
  {Ceperley}(1993)}]{velazco1993discussion}%
  \BibitemOpen
  \bibfield  {author} {\bibinfo {author} {\bibfnamefont {J.~E.}\ \bibnamefont
  {Velazco}}\ and\ \bibinfo {author} {\bibfnamefont {P.~H.}\ \bibnamefont
  {Ceperley}},\ }\bibfield  {title} {\bibinfo {title} {A discussion of rotating
  wave fields for microwave applications},\ }\href@noop {} {\bibfield
  {journal} {\bibinfo  {journal} {IEEE transactions on microwave theory and
  techniques}\ }\textbf {\bibinfo {volume} {41}},\ \bibinfo {pages} {330}
  (\bibinfo {year} {1993})}\BibitemShut {NoStop}%
\bibitem [{\citenamefont {Velazco}\ and\ \citenamefont
  {Ceperley}(2003)}]{velazco2003development}%
  \BibitemOpen
  \bibfield  {author} {\bibinfo {author} {\bibfnamefont {J.~E.}\ \bibnamefont
  {Velazco}}\ and\ \bibinfo {author} {\bibfnamefont {P.~H.}\ \bibnamefont
  {Ceperley}},\ }\bibfield  {title} {\bibinfo {title} {Development of a compact
  rotating-wave electron beam accelerator},\ }in\ \href@noop {} {\emph
  {\bibinfo {booktitle} {AIP Conference Proceedings}}},\ Vol.\ \bibinfo
  {volume} {680}\ (\bibinfo {organization} {American Institute of Physics},\
  \bibinfo {year} {2003})\ pp.\ \bibinfo {pages} {972--976}\BibitemShut
  {NoStop}%
\bibitem [{\citenamefont {Velazco}\ \emph {et~al.}(2016)\citenamefont
  {Velazco}, \citenamefont {Taylor}, \citenamefont {Liu}, \citenamefont
  {Hodyss},\ and\ \citenamefont {Allwood}}]{velazco2016novel}%
  \BibitemOpen
  \bibfield  {author} {\bibinfo {author} {\bibfnamefont {J.~E.}\ \bibnamefont
  {Velazco}}, \bibinfo {author} {\bibfnamefont {M.}~\bibnamefont {Taylor}},
  \bibinfo {author} {\bibfnamefont {Y.}~\bibnamefont {Liu}}, \bibinfo {author}
  {\bibfnamefont {R.}~\bibnamefont {Hodyss}},\ and\ \bibinfo {author}
  {\bibfnamefont {A.}~\bibnamefont {Allwood}},\ }\bibfield  {title} {\bibinfo
  {title} {A novel rotating-wave x-ray source for analysis of the martian
  landscape},\ }\href@noop {} {\bibfield  {journal} {\bibinfo  {journal} {The
  Interplanetary Network Progress Report}\ }\textbf {\bibinfo {volume} {42}},\
  \bibinfo {pages} {207} (\bibinfo {year} {2016})}\BibitemShut {NoStop}%
\bibitem [{\citenamefont {Orozco}\ \emph {et~al.}(2024)\citenamefont {Orozco},
  \citenamefont {Tsygankov}, \citenamefont {Parada-Becerra}, \citenamefont
  {Hern{\'a}ndez}, \citenamefont {Barrag{\'a}n},\ and\ \citenamefont
  {Martinez-Amariz}}]{orozco2024electron}%
  \BibitemOpen
  \bibfield  {author} {\bibinfo {author} {\bibfnamefont {E.~A.}\ \bibnamefont
  {Orozco}}, \bibinfo {author} {\bibfnamefont {P.}~\bibnamefont {Tsygankov}},
  \bibinfo {author} {\bibfnamefont {F.}~\bibnamefont {Parada-Becerra}},
  \bibinfo {author} {\bibfnamefont {A.}~\bibnamefont {Hern{\'a}ndez}}, \bibinfo
  {author} {\bibfnamefont {Y.}~\bibnamefont {Barrag{\'a}n}},\ and\ \bibinfo
  {author} {\bibfnamefont {A.}~\bibnamefont {Martinez-Amariz}},\ }\bibfield
  {title} {\bibinfo {title} {Electron nonlinear dynamics in a compact
  accelerator based on the circular rotating tm110 mode},\ }\href@noop {}
  {\bibfield  {journal} {\bibinfo  {journal} {Communications in Nonlinear
  Science and Numerical Simulation}\ }\textbf {\bibinfo {volume} {130}},\
  \bibinfo {pages} {107740} (\bibinfo {year} {2024})}\BibitemShut {NoStop}%
\bibitem [{\citenamefont {Golovanivsky}(1980)}]{golovanivsky1980autoresonant}%
  \BibitemOpen
  \bibfield  {author} {\bibinfo {author} {\bibfnamefont {K.}~\bibnamefont
  {Golovanivsky}},\ }\bibfield  {title} {\bibinfo {title} {Autoresonant
  acceleration of electrons at nonlinear ecr in a magnetic field which is
  smoothly growing in time},\ }\href@noop {} {\bibfield  {journal} {\bibinfo
  {journal} {Physica Scripta}\ }\textbf {\bibinfo {volume} {22}},\ \bibinfo
  {pages} {126} (\bibinfo {year} {1980})}\BibitemShut {NoStop}%
\bibitem [{\citenamefont {Golovanivsky}(1982)}]{golovanivsky1982gyrac}%
  \BibitemOpen
  \bibfield  {author} {\bibinfo {author} {\bibfnamefont {K.}~\bibnamefont
  {Golovanivsky}},\ }\bibfield  {title} {\bibinfo {title} {The gyrac: A
  proposed gyro-resonant accelerator of electrons},\ }\href@noop {} {\bibfield
  {journal} {\bibinfo  {journal} {IEEE Transactions on Plasma Science}\
  }\textbf {\bibinfo {volume} {10}},\ \bibinfo {pages} {120} (\bibinfo {year}
  {1982})}\BibitemShut {NoStop}%
\bibitem [{\citenamefont {Golovanivsky}(1983)}]{golovanivsky1983gyromagnetic}%
  \BibitemOpen
  \bibfield  {author} {\bibinfo {author} {\bibfnamefont {K.}~\bibnamefont
  {Golovanivsky}},\ }\bibfield  {title} {\bibinfo {title} {The gyromagnetic
  autoresonance},\ }\href@noop {} {\bibfield  {journal} {\bibinfo  {journal}
  {IEEE Transactions on plasma science}\ }\textbf {\bibinfo {volume} {11}},\
  \bibinfo {pages} {28} (\bibinfo {year} {1983})}\BibitemShut {NoStop}%
\bibitem [{\citenamefont {Gal}(1989)}]{gal1989gyrac}%
  \BibitemOpen
  \bibfield  {author} {\bibinfo {author} {\bibfnamefont {O.}~\bibnamefont
  {Gal}},\ }\bibfield  {title} {\bibinfo {title} {Gyrac: a compact, cyclic
  electron accelerator},\ }\href@noop {} {\bibfield  {journal} {\bibinfo
  {journal} {IEEE Transactions on Plasma Science}\ }\textbf {\bibinfo {volume}
  {17}},\ \bibinfo {pages} {622} (\bibinfo {year} {1989})}\BibitemShut
  {NoStop}%
\bibitem [{\citenamefont {Inoue}\ \emph {et~al.}(2014)\citenamefont {Inoue},
  \citenamefont {Hattori}, \citenamefont {Sugimoto},\ and\ \citenamefont
  {Sasai}}]{inoue2014design}%
  \BibitemOpen
  \bibfield  {author} {\bibinfo {author} {\bibfnamefont {T.}~\bibnamefont
  {Inoue}}, \bibinfo {author} {\bibfnamefont {T.}~\bibnamefont {Hattori}},
  \bibinfo {author} {\bibfnamefont {S.}~\bibnamefont {Sugimoto}},\ and\
  \bibinfo {author} {\bibfnamefont {K.}~\bibnamefont {Sasai}},\ }\bibfield
  {title} {\bibinfo {title} {Design study of electron cyclotron resonance-ion
  plasma accelerator for heavy ion cancer therapy},\ }\href@noop {} {\bibfield
  {journal} {\bibinfo  {journal} {Review of Scientific Instruments}\ }\textbf
  {\bibinfo {volume} {85}} (\bibinfo {year} {2014})}\BibitemShut {NoStop}%
\bibitem [{\citenamefont {Andreev}\ \emph {et~al.}(2017)\citenamefont
  {Andreev}, \citenamefont {Chuprov}, \citenamefont {Ilgisonis}, \citenamefont
  {Novitsky},\ and\ \citenamefont {Umnov}}]{andreev2017gyromagnetic}%
  \BibitemOpen
  \bibfield  {author} {\bibinfo {author} {\bibfnamefont {V.}~\bibnamefont
  {Andreev}}, \bibinfo {author} {\bibfnamefont {D.}~\bibnamefont {Chuprov}},
  \bibinfo {author} {\bibfnamefont {V.}~\bibnamefont {Ilgisonis}}, \bibinfo
  {author} {\bibfnamefont {A.}~\bibnamefont {Novitsky}},\ and\ \bibinfo
  {author} {\bibfnamefont {A.}~\bibnamefont {Umnov}},\ }\bibfield  {title}
  {\bibinfo {title} {Gyromagnetic autoresonance plasma bunches in a magnetic
  mirror},\ }\href@noop {} {\bibfield  {journal} {\bibinfo  {journal} {Physics
  of Plasmas}\ }\textbf {\bibinfo {volume} {24}} (\bibinfo {year}
  {2017})}\BibitemShut {NoStop}%
\bibitem [{\citenamefont {Andreev}\ \emph {et~al.}(2020)\citenamefont
  {Andreev}, \citenamefont {Ilgisonis}, \citenamefont {Novitsky},\ and\
  \citenamefont {Umnov}}]{andreev2020generation}%
  \BibitemOpen
  \bibfield  {author} {\bibinfo {author} {\bibfnamefont {V.}~\bibnamefont
  {Andreev}}, \bibinfo {author} {\bibfnamefont {V.}~\bibnamefont {Ilgisonis}},
  \bibinfo {author} {\bibfnamefont {A.}~\bibnamefont {Novitsky}},\ and\
  \bibinfo {author} {\bibfnamefont {A.}~\bibnamefont {Umnov}},\ }\bibfield
  {title} {\bibinfo {title} {Generation of plasma bunches under conditions of
  gyromagnetic autoresonance in a long magnetic mirror machine: Computational
  experiment},\ }\href@noop {} {\bibfield  {journal} {\bibinfo  {journal}
  {Plasma Physics Reports}\ }\textbf {\bibinfo {volume} {46}},\ \bibinfo
  {pages} {756} (\bibinfo {year} {2020})}\BibitemShut {NoStop}%
\bibitem [{\citenamefont {Andreev}\ \emph {et~al.}(2021)\citenamefont
  {Andreev}, \citenamefont {Novitsky},\ and\ \citenamefont
  {Umnov}}]{andreev2021autoresonance}%
  \BibitemOpen
  \bibfield  {author} {\bibinfo {author} {\bibfnamefont {V.}~\bibnamefont
  {Andreev}}, \bibinfo {author} {\bibfnamefont {A.}~\bibnamefont {Novitsky}},\
  and\ \bibinfo {author} {\bibfnamefont {A.}~\bibnamefont {Umnov}},\ }\bibfield
   {title} {\bibinfo {title} {Autoresonance phenomenon in a long mirror},\
  }\href@noop {} {\bibfield  {journal} {\bibinfo  {journal} {Physics of
  Plasmas}\ }\textbf {\bibinfo {volume} {28}} (\bibinfo {year}
  {2021})}\BibitemShut {NoStop}%
\bibitem [{\citenamefont {Joshi}\ and\ \citenamefont
  {Bhattacharjee}(2019)}]{joshi2019design}%
  \BibitemOpen
  \bibfield  {author} {\bibinfo {author} {\bibfnamefont {M.~K.}\ \bibnamefont
  {Joshi}}\ and\ \bibinfo {author} {\bibfnamefont {R.}~\bibnamefont
  {Bhattacharjee}},\ }\bibfield  {title} {\bibinfo {title} {Design of a
  rectangular waveguide to cylindrical cavity mode launcher for te011 mode with
  maximum quality-factor},\ }\href@noop {} {\bibfield  {journal} {\bibinfo
  {journal} {International Journal of RF and Microwave Computer-Aided
  Engineering}\ }\textbf {\bibinfo {volume} {29}},\ \bibinfo {pages} {e21825}
  (\bibinfo {year} {2019})}\BibitemShut {NoStop}%
\bibitem [{\citenamefont {Vennemann}(2015)}]{vennemann2015construction}%
  \BibitemOpen
  \bibfield  {author} {\bibinfo {author} {\bibfnamefont {T.~A.}\ \bibnamefont
  {Vennemann}},\ }\bibfield  {title} {\bibinfo {title} {Construction of a 4 ghz
  resonant cavity operating in te011 mode: Simulation and experiments},\
  }\href@noop {} {\  (\bibinfo {year} {2015})}\BibitemShut {NoStop}%
\bibitem [{\citenamefont {Guo}\ \emph {et~al.}(2018)\citenamefont {Guo},
  \citenamefont {Henry}, \citenamefont {Poelker}, \citenamefont {Rimmer},
  \citenamefont {Suleiman},\ and\ \citenamefont {Wang}}]{guo2018using}%
  \BibitemOpen
  \bibfield  {author} {\bibinfo {author} {\bibfnamefont {J.}~\bibnamefont
  {Guo}}, \bibinfo {author} {\bibfnamefont {J.~E.}\ \bibnamefont {Henry}},
  \bibinfo {author} {\bibfnamefont {M.}~\bibnamefont {Poelker}}, \bibinfo
  {author} {\bibfnamefont {R.~A.}\ \bibnamefont {Rimmer}}, \bibinfo {author}
  {\bibfnamefont {R.~S.}\ \bibnamefont {Suleiman}},\ and\ \bibinfo {author}
  {\bibfnamefont {H.}~\bibnamefont {Wang}},\ }\href@noop {} {\emph {\bibinfo
  {title} {Using a TE011 Cavity as a Magnetic Momentum Monitor}}},\ \bibinfo
  {type} {Tech. Rep.}\ (\bibinfo  {institution} {Thomas Jefferson National
  Accelerator Facility (TJNAF), Newport News, VA~…},\ \bibinfo {year}
  {2018})\BibitemShut {NoStop}%
\bibitem [{\citenamefont {Pozar}(2021)}]{pozar2021microwave}%
  \BibitemOpen
  \bibfield  {author} {\bibinfo {author} {\bibfnamefont {D.~M.}\ \bibnamefont
  {Pozar}},\ }\href@noop {} {\emph {\bibinfo {title} {Microwave engineering:
  theory and techniques}}}\ (\bibinfo  {publisher} {John wiley \& sons},\
  \bibinfo {year} {2021})\BibitemShut {NoStop}%
\bibitem [{\citenamefont {Birdsall}\ and\ \citenamefont
  {Langdon}(2004)}]{birdsall2004plasma}%
  \BibitemOpen
  \bibfield  {author} {\bibinfo {author} {\bibfnamefont {C.~K.}\ \bibnamefont
  {Birdsall}}\ and\ \bibinfo {author} {\bibfnamefont {A.~B.}\ \bibnamefont
  {Langdon}},\ }\href@noop {} {\emph {\bibinfo {title} {Plasma physics via
  computer simulation}}}\ (\bibinfo  {publisher} {CRC press},\ \bibinfo {year}
  {2004})\BibitemShut {NoStop}%
\bibitem [{\citenamefont {Qin}\ \emph {et~al.}(2013)\citenamefont {Qin},
  \citenamefont {Zhang}, \citenamefont {Xiao}, \citenamefont {Liu},
  \citenamefont {Sun},\ and\ \citenamefont {Tang}}]{qin2013boris}%
  \BibitemOpen
  \bibfield  {author} {\bibinfo {author} {\bibfnamefont {H.}~\bibnamefont
  {Qin}}, \bibinfo {author} {\bibfnamefont {S.}~\bibnamefont {Zhang}}, \bibinfo
  {author} {\bibfnamefont {J.}~\bibnamefont {Xiao}}, \bibinfo {author}
  {\bibfnamefont {J.}~\bibnamefont {Liu}}, \bibinfo {author} {\bibfnamefont
  {Y.}~\bibnamefont {Sun}},\ and\ \bibinfo {author} {\bibfnamefont {W.~M.}\
  \bibnamefont {Tang}},\ }\bibfield  {title} {\bibinfo {title} {Why is boris
  algorithm so good?},\ }\href@noop {} {\bibfield  {journal} {\bibinfo
  {journal} {Physics of Plasmas}\ }\textbf {\bibinfo {volume} {20}},\ \bibinfo
  {pages} {084503} (\bibinfo {year} {2013})}\BibitemShut {NoStop}%
\end{thebibliography}%

\end{document}